\documentclass[preprint,nofootinbib]{revtex4}
\usepackage{amsfonts}
\usepackage{amsmath}
\usepackage{amssymb}
\usepackage{graphicx}
\usepackage[usenames]{color}

\begin{document}

\title{Bulk Viscosity and Relaxation Time of Causal Dissipative Relativistic Fluid Dynamics}
\author{Xu-Guang Huang$^{1,2}$}
\author{Takeshi Kodama$^{3}$}
\author{Tomoi Koide$^1$}
\author{Dirk H. Rischke$^{1,2}$}
\affiliation{$^1$ Frankfurt Institute for Advanced Studies, D-60438 Frankfurt am Main,
Germany\\
$^2$ Institut f\"ur Theoretische Physik, J. W. Goethe-Universit\"at,D-60438
Frankfurt am Main, Germany\\
$^3$Instituto de F\'isica, Universidade Federal do Rio de Janeiro, C. P.
68528, 21945-970, Rio de Janeiro, Brasil}

\begin{abstract}
The microscopic formulae of the bulk viscosity $\zeta $ and the
corresponding relaxation time $\tau _{\Pi }$ in causal dissipative relativistic
fluid dynamics are derived by using the projection operator method. 
In applying these formulae to the pionic fluid, we find that the renormalizable energy-momentum tensor should be employed to obtain consistent results.
In the leading order approximation in the
chiral perturbation theory, the relaxation time is enhanced
near the QCD phase transition and $\tau _{\Pi }$ and $\zeta $ are related as
$\tau _{\Pi }=\zeta /[\beta \{(1/3-c_{s}^{2})(\varepsilon +P)-2(\varepsilon
-3P)/9\}]$, where $\varepsilon $, $P$ and $c_{s}$ are the energy density,
pressure and velocity of sound, respectively. The predicted $\zeta $ and $%
\tau _{\Pi }$ should satisfy the so-called causality condition. We compare
our result with the results of
the kinetic calculation by Israel and Stewart and the string theory,
and confirm that all the three approaches are consistent with the causality
condition. 
\end{abstract}

\maketitle


\section{introduction}

\label{sec:intro}

Relativistic hydrodynamics is an important tool to describe high
energy flow phenomena in different areas, such as heavy-ion collisions,
relativistic astrophysics and cosmology, although its theoretical
foundation has not yet been fully established, in particular, when
dissipative processes are involved \cite{koide_review}.
The simplest formulation is a
relativistic covariant extension of the non-relativistic Navier-Stokes
equation. It is, however, known that relativistic Navier-Stokes theory
suffers the problem of relativistic acausality and instability \cite%
{his,dkkm3,pu}.

After the seminal work by Israel and Stewart \cite{is}, so far many
different approaches of the relativistic hydrodynamics which are consistent
with causality have been proposed \cite{dkkm1,dkkm3,dkkm4, jou, pratt,
calzetta, con, else}.
In the following, we call these theories as the causal dissipative relativistic hydrodynamics (CDR) \cite{koide_review}.
The crucial difference of any CDR theories
from the Navier-Stokes theory can be characterized by the introduction of
finite relaxation times in the definitions of irreversible currents.
This aspect is somehow overlooked but has an important consequence. That is,
any relativistically causal fluids will be \textit{non-Newtonian} in the
sense that irreversible currents are no longer simply proportional to the
corresponding thermodynamic forces while their \textit{Newtonian}
counterparts are.

In hydrodynamics, all transport coefficients such as the shear viscosity and
the bulk viscosity are inputs and should be determined from a microscopic
theory. In the case of the Navier-Stokes theory, the coefficients are
usually calculated by using two different approaches. One is the kinetic
approach based mainly on the Boltzmann equation, and the other is the
microscopic approach using the Green-Kubo-Nakano (GKN) formula.

Strictly speaking, the kinetic approach is applicable only to
rarefied gas and not reliable in practice to calculate the transport
coefficients for finite density systems. For example, the density expansion
of the shear viscosity $\eta (\rho)$ in three dimensional space is given by
\cite{dorfman}
\begin{equation}
\eta (\rho )=\eta _{0}+\eta _{1}\rho +\eta _{2}\rho ^{2}\ln \rho +\cdots .
\end{equation}%
What we can calculate from the Boltzmann equation is only the first term $%
\eta _{0}$. This is because the Boltzmann equation is the lowest order
approximation of the BBGKY hierarchy equation and does not contain the
information of multiple collisions which violates the important assumption of
the Boltzmann equation, that is, the molecular chaos. To calculate the
further expansion coefficients $\eta _{1}$ and $\eta _{2}$ in the kinetic
approach, we have to use, for example, the Bogoliubov-Choh-Uhlenbeck
equation which is a generalized Boltzmann equation \cite{dorfman}. Unfortunately,
systematic generalization of the relativistic Boltzmann equation is not yet
known.

On the other hand, the GKN formula does not have such a limitation
with respect to the finiteness of density as far as we know because
it is derived from the microscopic theory, quantum field theory
\cite{physicstoday}.
However, for a relativistic fluid, we have to be careful because
the Newtonian property of a fluid is assumed to derive the GKN formula.
Thus new formulism is needed to calculate the transport coefficients of CDR.

Transport phenomena such as viscosities, diffusion and heat
conduction in hydrodynamics are related rather to the changes in a
sort of boundary conditions than to the responses of the system to
an external mechanical perturbation. For this reason, the required
formulation should be different from the ordinary linear response theory.
This fact is already emphasized by Kubo \cite{kubo}.
As a matter of fact, the well-known expressions of transport coefficients of
relativistic hydrodynamics are obtained by using the non-equilibrium
statistical operator method by Zubarev \cite{zuba,hosoya}.
So far, several different approaches have been proposed to
calculate the transport coefficients : 
indirect Kubo method, Langevin-Fokker-Planck method, regression hypothesis
based method, local equilibrium approach, external reservoir method, and
prediction theory, and so on. See Ref. \cite{review-zwanzig} for details.

Recently, the authors of the present paper proposed a new microscopic
formula to calculate the shear viscosity and the corresponding relaxation
time of CDR using the projection operator method,
which belongs to the Langevin-Fokker-Planck method in the classification mentioned
above \cite{knk}. Our formula is expressed in terms
of the time correlation functions of microscopic currents and is a natural
extension of the GKN formula. We showed that for the shear viscosity,
it reproduces the GKN results in the leading order. When applied to a
Navier-Stokes fluid and non-relativistic diffusion coefficients, our
approach reproduces the well-known results, as is discussed in Ref.
\cite{koide, koide2}.

The purpose of this paper is to derive the microscopic formulae of
the bulk viscosity $\zeta $ and the corresponding relaxation time $\tau
_{\Pi }$ in the framework of the projection operator method \cite{dhkr}.
We apply the result to the pionic fluid and calculate in the
leading order approximation in the chiral perturbation theory 
with the renormalizable energy-momentum tensor. We find that
the relaxation time $\tau _{\Pi }$ is enhanced around the temperature
near the QCD phase transition.

We further discuss the differences among our formalism,
the kinetic theory \cite{is,muronga,dirk,dkr} and the string theory \cite{st}.
In a CDR, the values of $\zeta $ and $\tau _{\Pi }$ should satisfy the causality
condition, which is derived by employing that the propagating speed of a physical
signal should not exceed the speed of light. If this condition is violated,
relativistic fluids becomes unstable \cite{dkkm3,pu}. We confirm that
the values of transport coefficients obtained by all the three
different approaches are consistent with the causality condition.

It should be noted that the bulk viscosity and the corresponding
relaxation time are important not only for heavy-ion collision physics, but
also for astrophysics, for example, the stability windows in parameter space
of rotating compact stars \cite{huang}, driving inflation in early universe
and the associating entropy production \cite{astro-ph/9601189}.

This paper is organized as follows. First we give a brief review of the
projection operator method in Sec.~\ref{sec:proj} for the sake of the later
convenience. In Sec.~\ref{sec:bulk}, we apply the method and derive
expressions for the bulk viscosity and the relaxation time. We calculate the
formulae in the leading order approximation and discuss the consistency with
the causality condition in Sec.~\ref{sec:leading}. Possible other
generalization of our result is discussed in Sec.~\ref{sec:higher}. The
summary of our results concludes this work in Sec.~\ref{sec:summary}.
Through this paper, we will use metric $g=\mathrm{diag}(+,-,-,-)$ and the
natural units $\hbar =c=k_{B}=1$.


\section{Projection Operator method}

\label{sec:proj}

For the later convenience, let us briefly review the projection operator
method ~\cite{zwanzig-book, reichl, pom_review}. It should be emphasized that the projection
operator method was firstly proposed by Nakajima \cite{nakajima}, although
it is often referred as the Mori-Zwanzig formalism due to the extensive use
and developments done by these authors \cite{zwanzig, mori}.

Many dynamical variables of practical interest, such as the
conserved quantities, usually vary slowly in time comparing to other
microscopic quantities. We call them gross variables. \footnote{
When a phase transition is present, in principle, we
should consider also the corresponding order parameters and soft
modes as the candidates of the gross (hydrodynamic) variables.
We do not, however, discuss such a case in this work \cite{chi_km}.}
In order to discuss the dynamics of these
slowly varying relevant variables, we need to introduce a coarse-graining
procedure to smoothing out the microscopic dynamics.
The projection operator method provides a systematic way to extract the information of
the relevant coarse-grained dynamics from underlining microscopic theories.

In the case of a quantum system, the full microscopic dynamics is described
by the Heisenberg equation of motion,
\begin{equation}
\partial _{t}O(t)=i[H,O(t)]\equiv iLO(t),
\end{equation}%
where $O$ is an arbitrary operator and $H$ is the Hamiltonian. For
simplicity, here we have assumed that $H$ is independent of time.
See Refs. \cite{shibata,km} for the case of a system with time-dependent
Hamiltonian. The second equality defines the Liouville operator $L$.
In order to project out the irrelevant information associated with variables of
microscopic (short) time scales, we introduce a time-independent projection
operator $P$ and its complementary operator $Q=1-P$, which satisfy the
following general properties,
\begin{equation}
P^{2}=P,~~~PQ=QP=0.
\end{equation}%
With the help of these operators, the Heisenberg equation of motion can be
re-expressed as \cite{mori,shibata,km}
\begin{eqnarray}
\frac{\partial }{\partial t}O(t) &=&e^{iLt}PiLO(0)+\int_{0}^{t}d\tau
e^{iL(t-\tau )}PiLQe^{iLQ\tau }iLO(0)  \notag  \label{tc} \\
&&+Qe^{iLQt}iLO(0).  \label{eqn:TC-1}
\end{eqnarray}

This is called the time-convolution (TC) equation and its r.~h.~s.
is composed of three distinct parts. The first term is called the
streaming term and usually corresponds to collective oscillations such as
plasma wave and spin wave. The second term is called the
memory term which turns into the dissipation term after
a coarse-graining procedure. The third term is
identified with the noise term after implementing coarse-graining of time,
as we will see later.
Thus this equation is considered as a generalized Langevin equation.
As a matter of fact, the memory term and the noise term
are related through the fluctuation-dissipation theorem of second kind \cite%
{mori,kk}. Note that the TC equation is very general and still
equivalent to the Heisenberg equation of motion.

The choice of the most appropriate projection operator depends on
the specific properties of a given system and also on the
coarse-graining procedure which we wish to introduce. If we
choose the projection operator so as to extract all the relevant gross (in
our case, hydrodynamic) variables, we can, in principle, derive hydrodynamic
equations from the TC equation. For this purpose, we use the Mori projection
operator \cite{mori}. Let
\begin{equation}
\mathbf{\bar{A}}=\left(
\begin{array}{c}
\bar{A}_{1} \\
\bar{A}_{2} \\
\vdots \\
\bar{A}_{n}%
\end{array}%
\right)  \label{pro-mori}
\end{equation}%
be a $n$-dimensional vector formed by $n$ time-independent
operators corresponding to the gross variables $\left\{ \bar{A}_{i}\right\}
,$ where the notation $\bar{A}$ is used to distinguish the
Schr\"{o}dinger operator from its Heisenberg form, $A=A\left( t\right)$.
We choose $\bar{A}=A\left( 0\right) $, so that $A\left(
t\right) = e^{iLt}\bar{A}$.

Then the time-independent Mori projection operator $P$ is defined
as \cite{mori}
\begin{equation}
P~O=\sum_{i=1,}^{n}c_{i}\bar{A}_{i}, \label{project}
\end{equation}%
where $O$ is an arbitrary operator, and the coefficient $c_{i}$ is given by
\begin{equation}
c_{i}=\sum_{j=1}^{n}(O,\bar{A}_{j}^{\dagger })\cdot (\mathbf{\bar{A}},%
\mathbf{\bar{A}}^{\dagger })_{ji}^{-1}.
\end{equation}%
Here $\left( X,Y\right)$ denotes the inner product of two
arbitrary operators $X$ and $Y$ (see below), and
$(\mathbf{\bar{A}},\mathbf{\bar{A}}^{\dagger })_{ji}^{-1}$
denotes $ji$ element of the inverse matrix of
$(\mathbf{\bar{A}},\mathbf{\bar{A}}^{\dagger }),$ i.e.,
\begin{equation}
\sum_{j}(\mathbf{\bar{A}},\mathbf{\bar{A}}^{\dagger })_{ij}^{-1}\cdot (\bar{A%
}_{j},\bar{A}_{k}^{\dagger })=\delta _{i,k}.
\end{equation}%
In this way, we expect that the relevant part of an arbitrary
operator is expressed as a function of the gross variables by operating
this projection operator.
In order to follow the dynamics of the gross
variables in time, we would need the time-dependent projections \cite{koide_t},
but for the present purpose of calculating the transport coefficients, the time independent
projection is sufficient.

We are still left with the freedom to choose the definition of the
inner product.
Here, following Ref. \cite{mori}, we use Kubo's canonical
correlation,
\begin{equation}
(X,Y)=\int_{0}^{\beta }\frac{d\lambda }{\beta }\mathrm{Tr}[\rho
_{eq}~e^{\lambda H}Xe^{-\lambda H}Y],
\end{equation}%
where $\rho _{eq}=e^{-\beta H}/\mathrm{Tr}[e^{-\beta H}]$ with $\beta $
being the inverse of temperature. One can see that, if it is a classical
system, Kubo's canonical correlation is reduced to the usual classical
thermal expectation value. Thus Kubo's canonical correlation is the quantum
generalization of the classical expectation values. It is easy to confirm
that
\begin{equation}
(iLX,Y)=-(X,iLY).
\end{equation}

Finally, the TC equation for the gross variable $\mathbf{%
A}\left( t\right) $ in the Heisenberg picture can be expressed as
\begin{equation}
\frac{\partial }{\partial t}\mathbf{A}(t)=i\Delta \ \mathbf{A}%
(t)-\int_{0}^{t}d\tau \mathbf{\Xi }(\tau )\mathbf{A}(t-\tau )+\mathbf{\xi }%
(t), \label{morieq}
\end{equation}%
where $\Delta $ and $\Xi $ are operators of $\left( n\times n\right) $%
matrices and $\mathbf{\xi }$ is a $n$-vector
whose elements are given by
\begin{eqnarray}
i\Delta _{ij} &=&\sum_{k}(iL\bar{A}_{i},\bar{A}_{k}^{\dagger })(\mathbf{\bar{%
A}},\mathbf{\bar{A}}^{\dagger })_{kj}^{-1}, \\
\mathbf{\Xi }_{ij}(t) &=&-\theta (t)\sum_{k}(iLQe^{iLQt}iL\bar{A}_{i},\bar{A}%
_{k}^{\dagger })(\mathbf{\bar{A}},\mathbf{\bar{A}}^{\dagger })_{kj}^{-1},
\label{xi} \\
{\xi }_{i}(t) &=&Qe^{iLQt}iL\bar{A}_{i}.
\end{eqnarray}

If the set of $n$-gross variables $\{\bar{A}_{i}\}$
is appropriately chosen so as to extract all the dynamics associated with
the slow hydrodynamic time scale, we expect that the dynamical variation
time-scale of the last term $\xi _{i}(t)$ of Eq.(\ref{morieq}) should be very small
compared to the hydrodynamic time scale, because the projection operator
$Q$ projects out components only orthogonal to
$\{\bar{A}_{i}\}$.
For this reason the term $\xi _{i}(t)$ is called the noise term.


\section{General formulae for bulk viscosity and relaxation time}

\label{sec:bulk}

In this section, we derive the microscopic formulae for the bulk viscosity $%
\zeta $ and the corresponding relaxation time $\tau _{\Pi }$. Our strategy
is as follows. We derive the evolution equation of the bulk viscous pressure
$\Pi $ from the TC equation, and compare the derived microscopic equation
with the phenomenological one to extract the microscopic formulae.
For our
purpose of obtaining the transport coefficient, it is sufficient to consider
small deviation from the stationary back-ground-fluid in thermal equilibrium.

The phenomenological equation of the bulk viscous pressure $\Pi $ in a CDR
is given by \cite{dkkm1,dkkm4}
\begin{equation}
\tau _{\Pi }u^{\mu }\partial _{\mu }\Pi +\Pi =-\zeta \partial _{\mu }u^{\mu
},  \label{eqn:phe_bulk1}
\end{equation}%
where $\zeta $, $\tau _{\Pi }$ and $u^{\mu }$ are the bulk viscosity, the
relaxation time and the fluid velocity, respectively. The first term on the
l.h.s. represents the retardation effect of $\Pi $ which is necessary to
satisfy relativistic causality. For $\tau _{\Pi }=0$, Eq.~(\ref%
{eqn:phe_bulk1}) is reduced to the usual Navier-Stokes constructive
equation.
In general, as is predicted from the kinetic theory,
it is possible to introduce more non-linear terms in Eq. (%
\ref{eqn:phe_bulk1}) but, as mentioned above, here we discuss only the
lowest order equation consistent with a CDR.

In order to avoid the possible influence from the shear stress tensor,
we consider a perturbation in infinite fluid in thermal equilibrium
having a planar symmetry in $\left( y,z\right)$ plane.
All the quantities associated with the perturbed fluid dynamics vary spatially only
along the $x$ direction. In this case, the fluid
velocity points to the $x$ direction (if one wants to discuss shear
viscosity only, one can choose the fluid velocity to point to the $x$
direction but vary spatially along the $y$ direction, as is done in Ref.
\cite{zwanzig-book,knk}). Then, the equation of continuity of the
energy-momentum tensor $T^{\mu \nu }$ in momentum space is given by
\begin{equation}
\partial _{t}T^{x0}(k^{x},t)=-ik_{x}T^{xx}(k^{x},t),  \label{txo}
\end{equation}%
where $k^{x}$ denotes the $x$-component of the
momentum vector ${\bf k}$.
On the other hand, Eq. (\ref{eqn:phe_bulk1}) is simplified as
\begin{equation}
\tau _{\Pi }\partial _{t}\Pi (k^{x},t)+\Pi (k^{x},t)=-\zeta
ik^{x}u^{x}(k^{x},t).  \label{pheno_bulk}
\end{equation}%
We use this equation as the definition of $\zeta $ and $\tau _{\Pi }$. Here $%
\Pi (k^{x},t),T^{x0}(k^{x},t)$ and $T^{xx}(k^{x},t)$ are the Fourier
transforms of $\Pi (x,t),T^{x0}(x,t)$ and $T^{xx}(x,t)$, respectively.

In order to obtain the microscopic expressions of
$\zeta $ and $\tau _{\Pi }$,
we derive the equation for $\Pi $ from the TC equation. For this purpose,
we have to choose appropriate gross variables included in
Eq. (\ref{pheno_bulk}) to define the projection operator.
Among $T^{x0},T^{xx},\Pi $ and $u^{x}$, the bulk viscous pressure
$\Pi $ is the deviation from the equilibrium pressure in the diagonal
components of the energy-momentum tensor. Thus we use the following operator
representation,
\begin{equation}
\Pi (k^{x})=-\frac{1}{3}T_{\mu }^{\mu }(k^{x})+\frac{1}{3}\langle T_{\mu
}^{\mu }(k^{x})\rangle _{eq},  \label{op_bulk}
\end{equation}%
where $\langle ~\cdots ~\rangle _{eq}$ represents the
equilibrium expectation value. Furthermore, as we will discuss later,
$u^{x} $ can be regarded as linearly dependent of $T^{0x}$
in the lowest order of the perturbation.
Then, Eq.~(\ref{pheno_bulk})
contains basically two independent gross variables,
\begin{equation}
\mathbf{\bar{A}}(k^{x})
=
\left(
\begin{array}{c}
\bar{T}^{0x}(k^{x})  \\
\bar{\Pi}(k^{x})
\end{array}
\right) ,
\end{equation}
where the bar notation, for example $\bar{\Pi},$ refers to the operator
value of $\Pi \left( t\right) $ at $t=0.$

According to Eq.~(\ref{project}), the projection operator $P$ is now defined
by
\begin{equation}
PO=\frac{(O,\bar{T}^{0x}(-k^{x}))}{(\bar{T}^{0x}(k^{x}),\bar{T}^{0x}(-k^{x}))%
}\bar{T}^{0x}(k^{x})+\frac{(O,\bar{\Pi}(-k^{x}))}{(\bar{\Pi}(k^{x}),\bar{\Pi}%
(-k^{x}))}\bar{\Pi}(k^{x}),
\end{equation}
Substituting it into Eq. (\ref{morieq}), we obtain the following two
equations,
\begin{eqnarray}
\partial _{t}T^{0x}(k^{x},t) &=&-ik^{x}\Pi (k^{x},t), \\
\partial _{t}\Pi (k^{x},t)
&=&-ik^{x}R_{k^{x}}T^{0x}(k^{x},t)-\int_{0}^{t}d\tau \Xi _{22}(k^{x},\tau
)\Pi (k^{x},t-\tau )+\xi (k^{x},t),  \label{langevin2}
\end{eqnarray}%
where
\begin{equation}
R_{k^{x}}=\frac{(\bar{\Pi}(k^{x}),\bar{\Pi}(-k^{x}))}{(\bar{T}^{0x}(k^{x}),%
\bar{T}^{0x}(-k^{x}))},
\end{equation}%
and we used
\begin{equation}
(iL\mathbf{\bar{A}},\mathbf{\bar{A}}^{\dagger })(\mathbf{\bar{A}},\mathbf{%
\bar{A}}^{\dagger })^{-1}=\left(
\begin{array}{cc}
0 & -ik^{x} \\
-ik^{x}R_{k^{x}} & 0%
\end{array}%
\right) .
\end{equation}%
Note that we consider homogeneous energy density and pressure. Then the
first equation is nothing but just Eq.(\ref{txo}), that is, the equation of
continuity, and the second equation describes the non-trivial evolution of
$\Pi $.

The exact expression of the memory function $\Xi _{22}$ is given in Ref.
\cite{kk}. However, if we are interested in the expression in the low $k^{x}$
limit, we can calculate it more easily. From Eq. (\ref{langevin2}), the
evolution of $(\Pi (k^{x},t),\bar{\Pi}(-k^{x}))$ is given by
\begin{equation}
\partial _{t}(\Pi (k^{x},t),\bar{\Pi}%
(-k^{x}))=-ik^{x}R_{k^{x}}^{B}(T^{0x}(k^{x},t),\bar{\Pi}(-k^{x}))-%
\int_{0}^{t}d\tau \Xi _{22}(k^{x},t-\tau )(\Pi (k^{x},\tau ),\bar{\Pi}%
(-k^{x})).
\end{equation}%
Here we used $(\xi (k^{x},t),\bar{\Pi}(-k^{x}))=0$, which is calculated from
the definition of $\xi(k^{x},t)$.
Then the Laplace transform of the memory function at low $k^{x}$
is
\begin{equation}
\Xi _{22}^{L}(k^{x},s)=\frac{1-sX^{L}(k^{x},s)}{X^{L}(k^{x},s)},
\end{equation}%
where
\begin{equation}
X^{L}(k^{x},s)\equiv \int_{0}^{\infty }dte^{-st}\frac{(\Pi (k^{x},t),\bar{\Pi%
}(-k^{x}))}{(\bar{\Pi}(k^{x}),\bar{\Pi}(-k^{x}))}.
\end{equation}

From the final value theorem of the Laplace transform, we can show that
\begin{eqnarray}
\lim_{s\rightarrow 0^{+}}sX^{L}(k^{x},s) &=&\lim_{t\rightarrow \infty }\frac{%
(\Pi (k^{x},t),\bar{\Pi}(-k^{x}))}{(\bar{\Pi}(k^{x}),\bar{\Pi}(-k^{x}))}
\notag \\
&=&\frac{\langle \Pi (k^{x},\infty )\rangle _{eq}\langle \bar{\Pi}%
(-k^{x})\rangle _{eq}}{(\bar{\Pi}(k^{x}),\bar{\Pi}(-k^{x}))}  \notag \\
&=&0.
\end{eqnarray}%
Here we used the mixing property of the ergodic theory. Finally, the memory
function in low $k^{x}$ and $s$ limit is given by
\begin{equation}
\Xi _{22}^{L}(k^{x},s)=\frac{1}{X^{L}(k^{x},s)}.
\end{equation}

To extract the phenomenological equation (\ref{pheno_bulk}),
we have to violate the time reversal symmetry.
For this purpose, we implement the coarse-graining of time.
Let us introduce a macroscopic time scale $\tau_M$ as follows
\begin{equation}
\tau_M = \epsilon t,
\end{equation}
where $\epsilon$ is a scale parameter and less than one.
Then the time-convolution integral is expressed as
\begin{equation}
\int_{0}^{\tau_M/\epsilon}d\tau \Xi _{22}(k^{x},\tau
)\Pi (k^{x},\tau_M/\epsilon-\tau ).
\end{equation}
When the microscopic and macroscopic time scales are clearly separated,
we can take the the vanishing $\epsilon$ limit.
Then the integral is given by
\begin{equation}
\lim_{\epsilon \rightarrow 0}\int_{0}^{\tau_M/\epsilon}d\tau \Xi _{22}(k^{x},\tau
)\Pi (k^{x},\tau_M/\epsilon-\tau )
=
\int_{0}^{\infty}d\tau \Xi _{22}(k^{x},\tau
)\Pi (k^{x},t) .
\end{equation}
We call this coarse-grainings the time-convolutionless (TCL)
approximation. Note that this approximation is very similar to the so-called
Markov approximation. In the present case, however,
there is still the memory effect for $\Pi $
even after the TCL approximation and we cannot call it the Markov approximation.
With this approximation, Eq. (\ref{langevin2}) is expressed as
\begin{eqnarray}
\partial _{t}\Pi (k^{x},t) &\approx
&-ik^{x}R_{k^{x}}T^{0x}(k^{x},t)-\int_{0}^{\infty }d\tau \Xi
_{22}(k^{x},\tau )~\Pi (k^{x},t)  \notag \\
&=&-ik^{x}R_{k^{x}}T^{0x}(k^{x},t)-\frac{1}{\tau ^{\Pi }(k^{x})}\Pi (k^{x},t)
\notag \\
&\approx &-ik^{x}R_{k^{x}}(\varepsilon +P)u^{x}(k^{x},t)-\frac{1}{\tau ^{\Pi
}(k^{x})}\Pi (k^{x},t).  \label{markov}
\end{eqnarray}%
Here, the noise term is neglected. The
function $\tau ^{\Pi }(k^{x})$ in the second line is defined by
\begin{equation}
\tau ^{\Pi }(k^{x})=X^{L}(k^{x},s=0).
\end{equation}
From the second line to the third line, we used the following
replacement
\begin{equation}
T^{0x}(k^{x},t)\simeq (\varepsilon +P)u^{x}(k^{x},t),
\end{equation}%
which comes from the expression of the phenomenological
energy-momentum tensor,
\begin{equation}
T^{0x}(x,t)=[\varepsilon +P+\Pi (x,t)]u^{x}(x,t),  \label{t_vel}
\end{equation}%
and is justified near the local rest frame.
Because we defined
the projection operator with $\Pi$ and $u^{x}$ by neglecting non-linear terms,
we cannot predict the coefficients of non-linear terms correctly
in the present calculation.
Thus, for the sake of consistency, we neglect the non-linear term
$\Pi (x,t) u^{x}(x,t)$ in Eq. (\ref{t_vel}).
The validity of the TCL approximation and the general comment for
the derivation of the non-linear term are discussed in Sec. \ref{sec:higher}.

By comparison Eq. (\ref{markov}) with Eq. (\ref{pheno_bulk}), we obtain the
following correspondences,
\begin{eqnarray}
\tau _{\Pi } &=& \lim_{s,\mathbf{k}\rightarrow
0}X^{L}(\mathbf{k},s), \\
\zeta &=&(\varepsilon +P)R_{\mathbf{0}}\lim_{s,\mathbf{k}\rightarrow
0}X^{L}(\mathbf{k},s).
\end{eqnarray}%
For the sake of convenience, we express these expressions in terms of the
retarded Green functions. Note that the correlation function $X^{L}(\mathbf{k%
},s)$ can be re-expressed as
\begin{eqnarray}
\lefteqn{X^{L}(\bf k,s)}  \notag \\
&=&-\frac{1}{\beta }\int_{0}^{\infty }dt\int d^{3}\mathbf{x}e^{-st-i\mathbf{%
k\cdot x}}\int_{t}^{\infty }d\tau \langle \Pi (\mathbf{x},\tau )\bar{\Pi}(%
\mathbf{0})\rangle _{ret}\left[ \int d^{3}\mathbf{x}_{1}e^{-i\mathbf{k\cdot
x_{1}}}(\bar{\Pi}(\mathbf{x}_{1}),\bar{\Pi}(\mathbf{0}))\right] ^{-1},
\notag \\
&&
\end{eqnarray}%
where the retarded Green function is defined by
\begin{eqnarray}
\langle \Pi (\mathbf{x},t)\Pi (\mathbf{x}_{1},\tau )\rangle _{ret}
&=&-i\theta (t-\tau )\mathrm{Tr}\{\rho _{eq}[\Pi (\mathbf{x},t),\Pi (\mathbf{%
x}_{1},\tau )]\}  \notag \\
&=&\int_{-\infty }^{\infty }\frac{d\omega d^{3}\mathbf{k}}{(2\pi )^{4}}%
G_{\Pi }^{R}(\omega ,\mathbf{k})e^{i\omega (t-\tau )}e^{-i\mathbf{k(x-x_{1})}%
}.
\end{eqnarray}
See Appendix \ref{app1} for details.

Then, finally, the formulae of $\zeta $ and $\tau _{\Pi }$ are given by
\begin{eqnarray}
\frac{\zeta }{\beta (\varepsilon +P)} &=&\frac{\zeta _{GKN}}{\beta ^{2}\int
d^{3}\mathbf{x}(\bar{T}^{0x}(\mathbf{x}),\bar{T}^{0x}(\mathbf{0}))}
\label{bulkc2} \\
&=&-\frac{i}{\beta }\frac{\displaystyle\lim_{\omega \rightarrow 0}\lim_{%
\mathbf{k}\rightarrow \mathbf{0}}\partial G_{\Pi }^{R}(\omega ,\mathbf{k}%
)/\partial \omega }{\displaystyle\lim_{\mathbf{k}\rightarrow \mathbf{0}%
}\lim_{\omega \rightarrow 0}G_{T^{0x}}^{R}(\omega ,\mathbf{k})},
\label{bulkc} \\
\frac{\tau _{\Pi }}{\beta } &=&\frac{\zeta _{GKN}}{\beta ^{2}\int d^{3}%
\mathbf{x}(\bar{\Pi}(\mathbf{x}),\bar{\Pi}(\mathbf{0}))}  \label{bulkt2} \\
&=&-\frac{i}{\beta }\frac{\displaystyle\lim_{\omega \rightarrow 0}\lim_{%
\mathbf{k}\rightarrow \mathbf{0}}\partial G_{\Pi }^{R}(\omega ,\mathbf{k}%
)/\partial \omega }{\displaystyle\lim_{\mathbf{k}\rightarrow \mathbf{0}%
}\lim_{\omega \rightarrow 0}G_{\Pi }^{R}(\omega ,\mathbf{k})}.  \label{bulkt}
\end{eqnarray}%
Here we have introduced the usual expression of the bulk viscosity for the
relativistic Navier-Stokes fluid in the GKN formula (obtained with the Zubarev method),
\begin{eqnarray}
\zeta _{GKN} &=&-\int d^{3}\mathbf{x}\int_{0}^{\infty }dt\int_{t}^{\infty
}d\tau \langle \Pi (\mathbf{x},\tau )\bar{\Pi}(\mathbf{0})\rangle _{ret}
\notag \\
&=&i\lim_{\omega \rightarrow 0}\lim_{\mathbf{k}\rightarrow \mathbf{0}}\frac{%
\partial G_{\Pi }^{R}(\omega ,\mathbf{k})}{\partial \omega },
\end{eqnarray}%
and one more retarded Green function,
\begin{equation}
G_{T^{0x}}^{R}(\omega ,\mathbf{k})=\int_{-\infty }^{\infty }dtd^{3}\mathbf{x}%
\langle T^{0x}(\mathbf{x},t)\bar{T}^{0x}(\mathbf{0})\rangle
_{ret}e^{-i\omega t}e^{i\mathbf{kx}}.
\end{equation}%
Equations (\ref{bulkc2})-(\ref{bulkt}) are our main results.
The bulk viscosity and it
relaxation time are expressed by the ratios of Green's functions and different
orderings of limits.


\section{Applications to hot pionic fluid}

\label{sec:leading}

As an application of our microscopic formulae (\ref{bulkc}) and (\ref{bulkt}%
), we will calculate the bulk viscosity $\zeta $ and relaxation time $\tau
_{\Pi }$ for hot pion fluid in confined phase
within an effective model.

Let $\phi $ be the scalar field for pions (we simply use a real scalar field to present pions since the charge does not affect the results). The usual definition of the
energy-momentum tensor of this field is
\begin{equation}
T^{\mu \nu }=\partial ^{\mu }\phi \partial ^{\nu }\phi -g^{\mu \nu }\mathcal{%
L},  \label{Tmunu}
\end{equation}%
where $\mathcal{L}$ is a Lagrangian density. In this case, the bulk
viscous pressure (\ref{op_bulk}) for non-interacting case,
would become
\begin{equation}
\Pi (\mathbf{x},t)=\frac{1}{3}[(\partial \phi (\mathbf{x},t))^{2}-2M^{2}\phi
^{2}(\mathbf{x},t)]-\frac{1}{3}\langle (\partial \phi (\mathbf{x}%
,t))^{2}-2M^{2}\phi ^{2}(\mathbf{x},t)\rangle _{eq},  \label{Pi_field}
\end{equation}%
where $M$ is the mass of pion.
However, these expressions are not
adequate for our purpose.
First, note that the above bulk viscous pressure
does not vanish even in the massless limit $M=0$, which does not
reflect the conformal property of the Lagrangian in this limit.
Furthermore, the energy-momentum tensor (\ref{Tmunu}) is not
renormalizable, \textit{i.e.}, its matrix elements depends directly
on the cut-off of the renormalized perturbation theory as is discussed in Ref. \cite{callan}.
Thus we introduce
the renormalizable energy-momentum tensor $\theta ^{\mu
\nu }$ following Ref. \cite{callan} as,
\begin{equation}
\theta ^{\mu \nu }(\mathbf{x},t)=T^{\mu \nu }(\mathbf{x},t)-\frac{1}{6}%
(\partial ^{\mu }\partial ^{\nu }-g^{\mu \nu }\partial ^{2})\phi ^{2}(%
\mathbf{x},t).  \label{Theta_munu}
\end{equation}%
Then the corresponding bulk viscous pressure for the non-interacting case is
given by
\begin{equation}
\Pi (\mathbf{x},t)=-\frac{1}{3}\theta _{\mu }^{\mu }(\mathbf{x},t)+\frac{1}{3%
}\langle \theta _{\mu }^{\mu }(\mathbf{x},t)\rangle _{eq}=-\frac{M^{2}}{3}%
(\phi ^{2}(\mathbf{x},t)-\langle \phi ^{2}(\mathbf{x},t)\rangle _{eq}),
\label{op_bulk2}
\end{equation}%
which recovers the conformal nature of the system in the vanishing limit of
$M$.
Note here that, for fermion field and gauge field, the usual definition of
energy-momentum tensor is already renormalizable and do not need any
re-definition of the energy-momentum tensor.

However, because of the reason which will be discussed in the end of this section,
this is still not the definition of the bulk viscous pressure which
is used in the following calculation.
We recall that the
behavior of the retarded Green function $G_{\Pi }^{R}$ in the low momentum
limit is not changed by adding an additional term which is proportional to
the energy density in the definition of the bulk viscous pressure.
Finally, we added an additional term which is proportional to the energy density
to define the bulk viscous pressure instead of Eq. (\ref{op_bulk}),
\begin{eqnarray}
\Pi (\mathbf{x},t) &=&-\frac{1}{3}\theta _{\mu }^{\mu }(\mathbf{x},t)+\left(
\frac{1}{3}-c_{s}^{2}\right) \theta ^{00}(\mathbf{x},t)+\left\langle \frac{1%
}{3}\theta _{\mu }^{\mu }(\mathbf{x},t)-\left( \frac{1}{3}-c_{s}^{2}\right)
\theta ^{00}(\mathbf{x},t)\right\rangle _{eq}  \notag \\
&=&-\frac{M^{2}}{3}\phi ^{2}(\mathbf{x},t)+\left( \frac{1}{3}%
-c_{s}^{2}\right) \theta ^{00}(\mathbf{x},t)+\left\langle \frac{M^{2}}{3}%
\phi ^{2}(\mathbf{x},t)-\left( \frac{1}{3}-c_{s}^{2}\right) \theta ^{00}(%
\mathbf{x},t)\right\rangle _{eq},  \label{Pi_New}
\end{eqnarray}%
where $c_{s}$ is the velocity of sound. This is the same definition of the
bulk viscous pressure discussed in Refs. \cite{joen,zuba}.
One can easily
see that this bulk viscous pressure still vanishes in the massless limit.

As the lightest particles, pions dominate the transport properties of QCD in
hadronic phase. From Eqs.~(\ref{bulkc2}) and (\ref{bulkt2}), once the
leading order result of $\zeta_{GKN}$ is obtained,
the corresponding leading order results
for $\zeta$ and $\tau_\Pi$ are obtained by substituting the
denominators on r.h.s by their non-interacting counterparts. A
straightforward calculation leads to
\begin{eqnarray}
\lim_{\mathbf{k \rightarrow 0}} \lim_{\omega \rightarrow 0} G^R_{T^{0x}}
(\omega ,\mathbf{k}) &=& -\varepsilon - P ,  \label{t0xt0x} \\
\lim_{\mathbf{k \rightarrow 0}} \lim_{\omega \rightarrow 0} G^R_{\Pi}
(\omega ,\mathbf{k}) &=& -\left( \frac{1}{3} - c^2_s \right) (\varepsilon +
P) + 2 \frac{\varepsilon - 3P}{9} ,  \label{pipi}
\end{eqnarray}
where $E_{\mathbf{p}} = \sqrt{\mathbf{p}^2 + M^2}$. The energy density and
pressure of the free pion gas are, respectively, given by
\begin{eqnarray}
\varepsilon &=& \frac{N_\pi}{V}\sum_{\mathbf{p}} E_{\mathbf{p}}f(E_{\mathbf{p%
}}), \\
P &=& \frac{N_\pi}{3V}\sum_{\mathbf{p}} \frac{\mathbf{p}^2}{E_{\mathbf{p}}}
f(E_{\mathbf{p}}),
\end{eqnarray}
where $f(x)$ is the Bose-Einstein distribution function $1/(e^{\beta x} - 1)$
and the prefactor $N_\pi=3$ counts the degeneracy of $\pi^+, \pi^-$ and $%
\pi^0$.

The r.~h.~s of Eq.(\ref{pipi}) has a term which contains ultraviolet divergent vacuum term. In order to obtain the finite result, we renormalized this vacuum term.  However, it should be noted that the renormalization
of this divergence is not trivial as is discussed in Ref. \cite{romson}.

In short, in the leading order approximation, the bulk viscosity and the
relaxation time are given by \cite{dhkr}
\begin{eqnarray}
\frac{\zeta}{\beta (\varepsilon + P)} &=& \frac{\zeta_{GKN}}{\beta
(\varepsilon + P)} ,  \label{lead_zeta} \\
\frac{\tau_{\Pi}}{\beta} &=& \frac{ \zeta_{GKN}}{\displaystyle \beta \left[
\left( \frac{1}{3} - c^2_s \right) (\varepsilon + P) - 2 \frac{\varepsilon -
3P}{9} \right]} .  \label{lead_tau}
\end{eqnarray}
The first equation shows that the bulk viscosity $\zeta$ is reduced to the
GKN bulk viscosity $\zeta_{GKN}$, similarly to the case of the shear
viscosity in the leading order calculation \cite{knk}. There already exist
several calculations for $\zeta_{GKN}$ \cite{various_bulk}. Thus we will not
discuss its behavior here.

\begin{figure}[t]
\includegraphics[scale=0.5]{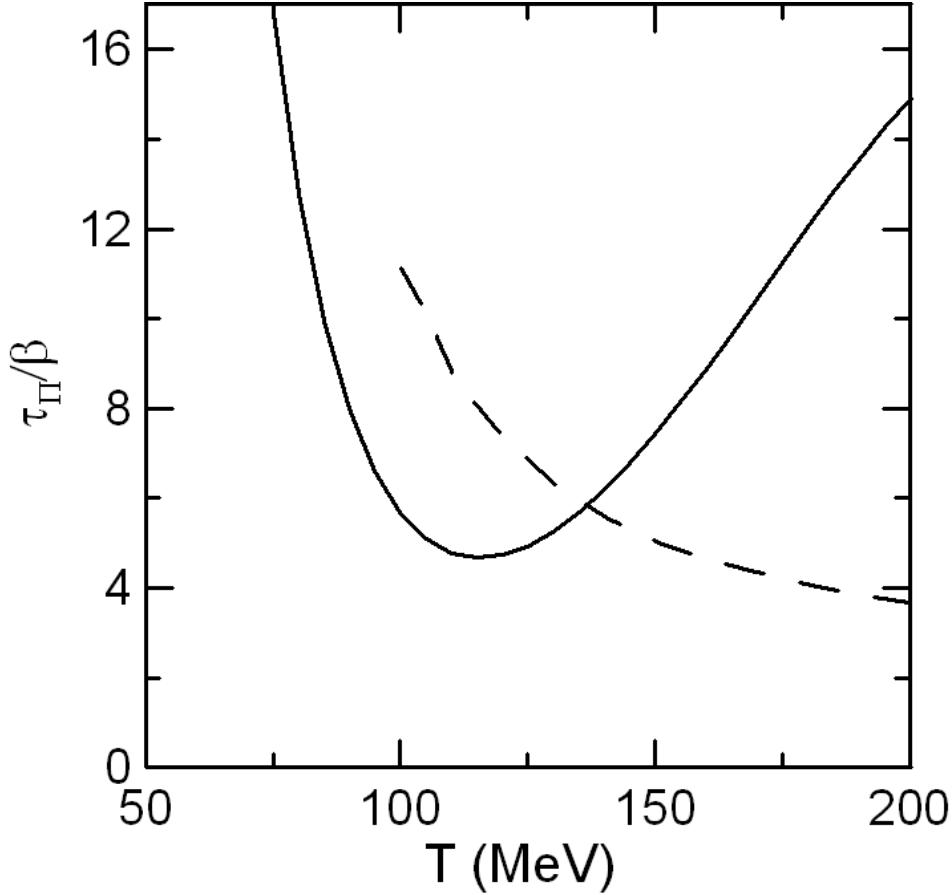}
\caption{The temperature dependence of the relaxation time of the bulk
viscous pressure $\protect\tau_{\Pi}/\protect\beta$ of the hadron phase. The
solid line represents the result of the leading order approximation in the
chiral perturbation theory. For the sake of comparison, the result of the 14
moment approximation from Ref. \protect\cite{ppv} is shown by the dashed
line. }
\label{fig:relaxt}
\end{figure}

By adopting the result of $\zeta_{GKN}$ calculated in the chiral
perturbation theory \cite{daniel}, we plot the temperature dependence of the
dimensionless ratio $\tau_{\Pi}/\beta$ in the hadron phase in Fig. \ref{fig:relaxt}.
For the sake of comparison,
the relaxation time calculated from the Boltzmann equation with Grad's moment method
is shown by the dashed line \cite{ppv}.
The order of the magnitude is same as the relaxation time of
the shear viscosity, $\tau_\pi/\beta$, which is shown in Fig. 1 in Ref. \cite%
{knk}. However, the temperature dependence of $\tau_{\Pi}/\beta$ is
non-trivial. As is shown in Ref. \cite{knk}, $\tau_\pi/\beta$ is a
monotonically decreasing function of temperature in the hadronic phase. On
the other case, $\tau_{\Pi}/\beta$, which is a decreasing function at low
temperature, starts to increase around $T=100$ MeV and shows maximum near
the QCD phase transition. This comes from the enhancement of $\zeta_{GKN}$
caused by the trace anomaly. See Ref. \cite{daniel} for details.

Now we compare our result with the results from Grad's method with the 14
moment approximation \cite{is,muronga} and the string theory \cite{st}. For
this purpose, it is convenient to consider the $\zeta$-$\tau_{\Pi}$ ratio,
because this quantity is independent of the choice of the collision term in
the Boltzmann equation. In our leading order result of pion, this ratio is
given by
\begin{eqnarray}
\frac{\zeta}{\tau_{\Pi}(\varepsilon + P)} = R_{\mathbf{0}} = \frac{\left(
\frac{1}{3} - c^2_s \right) (\varepsilon + P) - \frac{2}{9} (\varepsilon -
3P)} {\displaystyle(\varepsilon + P)}.  \label{eqn:zeta_tau}
\end{eqnarray}
In the result of the string theory, this ratio is given by \cite{st}
\begin{equation}
\frac{\zeta}{\tau_{\Pi}(\varepsilon + P)} = \frac{(1/3 - c^2_s)}{2-\ln 2}.
\end{equation}
In the 14 moment approximation,
this ratio is calculated by using a function $\beta_0$
which is defined by Eq. (7.8c) of Ref. \cite{is}.

\begin{figure}[!htb]
\includegraphics[scale=0.5]{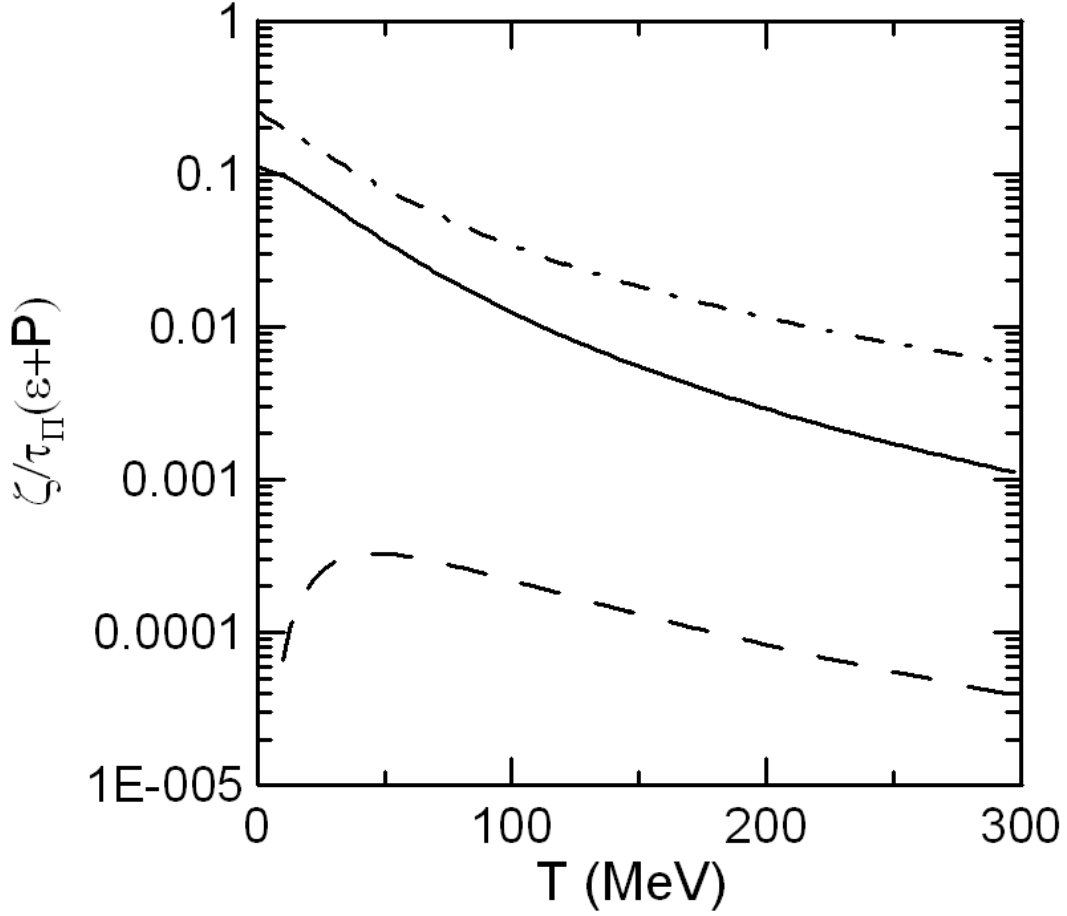}
\caption{The temperature dependence of the $\protect\zeta$-$\protect\tau%
_{\Pi}$ ratio. The dot-dashed, solid and dashed lines represents the results
of the 14 moment approximation, our formula (\protect\ref{eqn:zeta_tau}) and
the string theory, respectively. }
\label{fig:zeta_tau}
\end{figure}

The temperature dependence of the $\zeta$-$\tau_{\Pi}$ ratio for pion is
shown in Fig. \ref{fig:zeta_tau}. The solid, dashed and dot-dashed lines
represent our microscopic formula (\ref{eqn:zeta_tau}),
the 14 moment approximation, and the string theory, respectively.
The string theory predicts the largest value of the ratio,
meanwhile the 14 moment approximation
estimates the smallest value. At high temperature, the three lines
are monotonically decreasing functions of temperature.
The qualitative difference is
observed at low temperate. The ratio of our formula and the string theory are
still finite at $T=0$ but that of Grad's method vanishes.
The meaning of this difference is related to quantum fluctuation which is not
included in the Boltzmann equation.
This result is reported in another paper comparing our result
with a new kinetic calculation based on the Boltzmann equation \cite{dhkr,dkr}.

It is worth for mentioning that this ratio
is closely related to the propagation velocity of signals
in CDR, which is given by \cite{dkkm3,dkkm4}
\begin{equation}
v_{g}=\sqrt{c_{s}^{2}+\frac{\zeta }{\tau _{\Pi }(\varepsilon +P)}}.
\end{equation}%
Here, we neglected the contribution from the shear stress tensor. As is
discussed in Refs. \cite{dkkm3,pu}, for hydrodynamics being causal and
stable, this group velocity should be smaller than the speed of light. Thus
this ratio should satisfy the following constraint,
\begin{equation}
\frac{\zeta }{\tau _{\Pi }(\varepsilon +P)}\leq 1-c_{s}^{2}.
\end{equation}%
This is the so-called causality condition \cite{dkkm1,dkkm3,pu}. One can
easily see from Fig. \ref{fig:zeta_tau} that all the three calculations
satisfy the causality condition.

As was mentioned before, we have used Eq.(\ref{Pi_New})
instead of Eq.(\ref{op_bulk2}) as the definition of the bulk viscous pressure.
Then we pointed out that both definitions should give the same result.
However, this is not trivial when there is a UV divergence.
As a matter of fact, the $\zeta/\tau_\Pi$ ratio calculate with Eq. (\ref{op_bulk2})
is different from the solid line in Fig. \ref{fig:zeta_tau} when we adapt the simple subtraction of the vacuum term as renormalization.
In this calculation, we believe the calculation with Eq.(\ref{Pi_New})
is more reliable than that with Eq.(\ref{op_bulk2}) because of the following three reasons.
1) When we use Eq.(\ref{op_bulk2}) and employ the simple renormalization,
the calculated $\zeta/\tau_\Pi$ violate the causality condition.
2) As is pointed out in Ref. \cite{joen}, if Eq.(\ref{op_bulk2}) is used,
the perturbative calculations collapses because of divergence. This problem is solved by
using Eq.(\ref{Pi_New}).
3) We can show that the $\zeta/\tau_\Pi$ calculated with Eq.(\ref{Pi_New}) is
consistent with the result from a new kinetic calculation,
as is discussed in Ref. \cite{dhkr}.

\section{Other possible generalization}

\label{sec:higher}

\subsection{Another approximation to the memory function}

In the derivation of Eq. (\ref{markov}), we replaced the
time-convolution integral of the memory term with the time-convolutionless
integral by assuming that the macroscopic time scale is clearly separated from the microscopic one.
On the other hand, this coarse-grainings may be formulated
as a following expansion of the memory term,
\begin{equation}
\int^{t}_0 d\tau \Xi_{22} (k_x,\tau) \Pi(k_x,t-\tau) = \int^{t}_0 d\tau
\Xi_{22} (k_x,\tau) \left[ \Pi(k_x,t) - \tau \frac{\partial}{\partial t}%
\Pi(k_x,t) + \cdots \right].
\end{equation}

When the higher order correction becomes important, the evolution equation
of the bulk viscous pressure is modified as
\begin{eqnarray}
&& \partial_t \Pi (k_x,t) \approx -ik_x R^B_{k_{x}} (\varepsilon + P) u^x
(k_x,t) -\int^\infty_0 d\tau \Xi_{22} (k_x,\tau) \left[ \Pi(k_x,t) - \tau
\frac{\partial}{\partial t}\Pi(k_x,t) \right]  \notag \\
&& \longrightarrow \tilde{\tau}_{\Pi}\partial_t \Pi (k_x,t) + \Pi(k_x,t) =
-ik_x \zeta u^x( k_x , t ).
\end{eqnarray}
Here, we replaced the upper limit of the integral by $\infty$, assuming that 
the dominant contribution of the memory function still comes from $\tau = 0$.
The expression of $\zeta$ is not changed by this correction. However the
relaxation time is modified by $\tilde{\tau}_{\Pi}$ which is given by
\begin{equation}
\tilde{\tau}_{\Pi} = \tau_{\Pi}\left( 1 + \lim_{s,\mathbf{k} \rightarrow 0}
\frac{\partial \Xi^L(\mathbf{k},s)}{\partial s} \right) .
\end{equation}
This can be expressed in terms of the retarded Green function as follows,
\begin{eqnarray}
\tilde{\tau_{\Pi}} &=& \tau_{\Pi}
\frac{%
\displaystyle (\lim_{\omega,\mathbf{k}\rightarrow 0} \partial^2
G^R_{\Pi}(\omega,\mathbf{k})/\partial \omega^2)( \lim_{\mathbf{k},\omega \rightarrow 0}
G^R_{\Pi}(\omega,\mathbf{k}))} {\displaystyle 2(
\lim_{\omega,\mathbf{k}\rightarrow 0} \partial G^R_{\Pi}(\omega,\mathbf{k}%
)/\partial \omega)^2 }  . \label{til_tau}
\end{eqnarray}

For example, let us assume an exponential form for the memory function,
\begin{equation}
\Xi_{22}({\bf 0},t) = A\Gamma e^{-\Gamma t},
\end{equation}
where $A$ and $\Gamma$ are parameters.
Then we obtain that
\begin{equation}
\tilde{\tau_{\Pi}} = \tau_{\Pi}
\left(
1 - \frac{A}{\Gamma}
\right).
\end{equation}
The result with the TCL approximation is reproduced in the infinite $\Gamma$ limit.

However, because of the following reasons, we consider that the $\tau_\Pi$ is more reliable than $\tilde{\tau}_\Pi$.
Let us consider the following equation,
\begin{equation}
\partial_t J(t) = F(t) + \int^t_0 d\tau \Xi(t-\tau) J(\tau). \label{model}
\end{equation}
From the initial and final value theorem of the Laplace transform, we can calculate the
initial and final values of $J(t)$ as follows,
\begin{eqnarray}
\lim_{t\rightarrow 0} J(t) &=& \lim_{s\rightarrow \infty} sJ^L(s) = \lim_{s\rightarrow \infty} s \frac{J (t=0) - F^L(s)}{s - \Xi^L(s)} = J (t=0) - F^L(\infty), \label{exact_exam}\\
\lim_{t\rightarrow \infty} J(t) &=& \lim_{s\rightarrow 0} sJ^L(s) = 0.
\end{eqnarray}
Here we assumed that $\Xi^L(0)$ is finite
because of the existence of the finite relaxation time.
Next, we approximate the time-convolution integral of the above equation
by using the Taylor expansion,
\begin{equation}
\partial_t J(t) = F(t) + A J(t) + B \partial_t J(t).
\end{equation}
Then the initial and final values are
\begin{eqnarray}
\lim_{t\rightarrow 0} J(t) &=& \frac{J (t=0) - BJ^L(t=0) - F^L(\infty)}{1-B}, \\
\lim_{t\rightarrow \infty} J(t) &=& \lim_{s\rightarrow 0} sJ^L(s) = 0.
\end{eqnarray}
In order to reproduce Eq. (\ref{exact_exam}), we have to set $B=0$.

As another reason, we consider the calculation of the transport coefficients of the
diffusion equation.
As was shown by Kadanoff and Martin \cite{kadanoff}
and later by one of the present authors \cite{koide_dif},
the ratio of the diffusion coefficient $D$ and the corresponding relaxation time $\tau_D$ is
exactly determined from a sum rule. See Eqs. (B.10), (B.16) and (B.28) of Ref. \cite{koide_dif} .
Then we obtain
\begin{equation}
\frac{D}{\tau_D}
= \frac{\int d^3{\bf x}({\bf J}({\bf x}),{\bf J}({\bf 0}))}{\int d^3{\bf x}(n({\bf x}),n({\bf 0}))},
\end{equation}
where $n$ is the conserved number density and $\partial_t n + \nabla {\bf J} = 0$.
In the projection operator method, this result is reproduced only when the TCL approximation is applied.

That is,
if we regard that the TCL approximation as the lowest order of the Taylor expansion and the next order correction is considered,
we obtain the result which is inconsistent with the initial value theorem and the sum rule.

When we observe with the time scale where
the structure of the memory function in Eq. (\ref{model}) cannot be neglected,
it means that there still exists macroscopic degrees of freedom in the memory function.
In order to finish the program of coarse-graining in the projection operator method,
we have to re-define the projection operator so as to extract
this macroscopic degree of freedom.
Then the form of hydrodynamics itself is changed and the obtained equation is
not given by Eq. (\ref{eqn:phe_bulk1}) anymore.
In other words,
once we assume that the hydrodynamic equation is given by Eq. (\ref{eqn:phe_bulk1}),
we must observe with the time scale where the structure of the memory function is negligible,
and hence, we must use the TCL approximation instead of the Taylor expansion.

As an example of the time dependence of the memory function, see Fig. 4
in Ref. \cite{chi_km}.
One can see that the memory function has finite values only around $t=0$.

\subsection{Non-linear terms}

In this work, we used the phenomenological equation (\ref{eqn:phe_bulk1}) as
the definition of the bulk viscosity $\zeta$ and the relaxation time $%
\tau_{\Pi}$. On the other hand, it is possible to introduce non-linear
terms to the phenomenological equation. For example, it might be possible to
derive the hydrodynamic equation for the rarefied gas from the Boltzmann
equation, by using Grad's moment method with the 14 moment approximation.
In particular, the equation of the bulk
viscous pressure is given by, \cite{muronga,dirk}
\begin{eqnarray}
\tau_{\Pi} \frac{d}{d\tau} \Pi + \Pi = - \zeta \partial_{\mu} u^{\mu} + a_1
\Pi \partial_{\mu}u^{\mu} + a_2 \nu^{\mu}\frac{d u_{\mu}}{d\tau } + a_3
\partial_{\mu} \nu^{\mu} + a_4 \nu^{\mu} \Delta_{\mu\nu} \partial^{\nu}
\alpha + a_5 \pi_{\mu\nu} \Delta^{\mu\nu\alpha\beta}
\partial_{\alpha}u_{\beta},  \notag \\
\end{eqnarray}
where
\begin{eqnarray}
\Delta^{\mu\nu} &=& g^{\mu\nu} -u^{\mu}u^{\nu}, \\
\Delta^{\mu\nu\alpha\beta} &=& \frac{1}{2} \left( \Delta^{\mu
\alpha}\Delta^{\nu\beta} + \Delta^{\mu \alpha}\Delta^{\nu\beta} - \frac{2}{3}
\Delta^{\mu \nu}\Delta^{\alpha \beta} \right) .
\end{eqnarray}
In order to obtain the microscopic expressions of these $a_i$'s in the
projection operator method, the definition of the projection operator must
be generalized so as to collect all gross variables appears in this
non-linear equation. This will be put in future work.


\section{Concluding remarks}

\label{sec:summary}

In this paper, we derived the microscopic formulae of the bulk viscosity $%
\zeta$ and the corresponding relaxation time $\tau_{\Pi}$ in causal
dissipative relativistic fluid dynamics
by using the projection operator method. Applying
these formulae to the pionic fluid and calculating in the leading order
approximation in the chiral perturbation theory, we found that the
relaxation time is enhanced around the temperature near the QCD phase transition,
and there is a simple relation $\tau_{\Pi} = \zeta/[\beta \{(1/3-c_s^2)(\varepsilon +
P)-2(\varepsilon - 3 P)/9\}]$ between $\zeta$ and $\tau_\Pi$.
We compared our result with the results of
Grad's moment method with the 14 moment approximation and the string theory by
calculating the $\zeta$-$\tau_{\Pi}$ ratio which is independent of the
choice of the collision term of the Boltzmann equation.
The ratio must be smaller than $(1-c^2_s)(\varepsilon+P)$ to
satisfy the causality condition. We confirmed that all the three approaches
are consistent with the causality condition.
Finally, we discussed that
the time-convolutionless approximation, which is used to derive the transport coefficients,
is consistent with exact results and we should not consider corrections to this approximation.

It should be emphasized that Grad's moment method with the 14 moment approximation
is not the unique method to
calculate the transport coefficients of CDR
from the Boltzmann equation.
Recently a new calculation method based on the Boltzmann equation was developed
and it was found that the calculated transport coefficients are different from those of
the 14 moment approximation \cite{dkr}.
These new results are completely consistent with the
leading order results of our formulae. This result was reported in
another paper \cite{dhkr}.

We thank T. Brauner, G. S. Denicol, D. Fernandez-Fraile, J. Noronha, H. Warringa, and Z.
Xu for useful discussions and comments. This work was (financially)
supported by FAPERJ, CNPq and the Helmholtz International Center for FAIR
within the framework of the LOEWE program (Landesoffensive zur Entwicklung
Wissenschaftlich- Okonomischer Exzellenz) launched by the State of Hesse.

\appendix 


\section{Relation between Kubo's canonical correlation and retarded Green's
function}

\label{app1}
In this appendix, we give a relation connecting the Kubo's canonical
correlation and the retarded Green's function. This relation is used to
obtain Eqs.~(\ref{bulkc}) and (\ref{bulkt}). From the definition of Kubo's
canonical correlation, we can transform as
\begin{eqnarray}
\left(\Pi(t,\mathbf{x}),\Pi(t^{\prime },\mathbf{x}^{\prime
})\right)&\equiv&\int_0^\beta\frac{d\lambda}{\beta}\mathrm{Tr}\left[%
\rho_{eq} e^{\lambda H}\Pi(t,\mathbf{x})e^{-\lambda H}\Pi(t^{\prime },%
\mathbf{x}^{\prime })\right]  \notag \\
&=& \int_0^\beta\frac{d\lambda}{\beta}\langle\Pi(t-i\lambda,\mathbf{x}%
)\Pi(t^{\prime },\mathbf{x}^{\prime })\rangle_{eq}  \notag \\
&=&-\int_0^\beta\frac{d\lambda}{\beta}\int_0^\infty ds\langle\Pi(t-t^{\prime
}-i\lambda,\mathbf{x})\frac{d}{ds}\Pi(s,\mathbf{x}^{\prime })\rangle_{eq}
\notag \\
&=&i\int_0^\infty ds\int_0^\beta\frac{d\lambda}{\beta}\frac{d}{d\lambda}%
\langle\Pi(t-t^{\prime }-i\lambda,\mathbf{x})\Pi(s,\mathbf{x}^{\prime
})\rangle_{eq}  \notag \\
&=&i\int_0^\infty \frac{ds}{\beta}\langle\left[\Pi(s,\mathbf{x}^{\prime
}),\Pi(t-t^{\prime },\mathbf{x})\right]\rangle_{eq}  \notag \\
&=& i\int_0^\infty \frac{ds}{\beta}\langle\left[\Pi(s,\mathbf{x}^{\prime
}),\Pi(t-t^{\prime },\mathbf{x})\right]\rangle_{eq}.
\end{eqnarray}
Furthermore, when $t=t^{\prime }$, we have
\begin{eqnarray}
\left(\Pi(0,\mathbf{x}),\Pi(0,\mathbf{x}^{\prime
})\right)&=&i\int_{-\infty}^\infty \frac{ds}{\beta}\theta(s)\langle\left[%
\Pi(s,\mathbf{x}^{\prime }),\Pi(0,\mathbf{x})\right]\rangle_{eq}  \notag \\
&=&-\int_{-\infty}^\infty \frac{ds}{\beta}G^R_\Pi(s,\mathbf{x}^{\prime }-%
\mathbf{x})  \notag \\
&=&-\frac{1}{\beta}\lim_{\omega\rightarrow0}G^R_\Pi(\omega,\mathbf{x}%
^{\prime }-\mathbf{x}).
\end{eqnarray}
Then we obtain
\begin{eqnarray}
\int d^3\mathbf{x}\left(\Pi(0,\mathbf{x}),\Pi(0,\mathbf{0})\right)&=&-\frac{1%
}{\beta}\lim_{\mathbf{k}\rightarrow\mathbf{0}}\lim_{\omega\rightarrow0}G^R_%
\Pi(\omega,\mathbf{k}).
\end{eqnarray}

\end{document}